\documentclass[aps,prb,amsmath,twocolumn,superscriptaddress,floatfix,footinbib,longbibliography]{revtex4}

\usepackage{graphicx}
\usepackage{epstopdf}
\usepackage{comment}
\usepackage[T1]{fontenc}
\usepackage[applemac]{inputenc}
\usepackage{lmodern}
\usepackage[english]{babel}

\usepackage{ae}
\usepackage{units}

\usepackage{amsmath,amssymb,natbib,bm}
\usepackage{psfrag}
\usepackage{subfigure}
\usepackage{amsthm}

\usepackage{fixmath}
\usepackage{booktabs}

\usepackage{slashed}

\usepackage[americaninductors]{circuitikz}
\usepackage{tikz}
\usetikzlibrary{arrows}

\usepackage{url}

\usepackage[colorlinks]{hyperref}

\usepackage{tabularx}

\hypersetup{%
 plainpages=true,
 breaklinks=true,
 hypertexnames=false,
 pageanchor=true,
 colorlinks=true,
 linkcolor={blue},
 citecolor={magenta},
 urlcolor={blue},
 anchorcolor={black}
 }

\renewcommand{\eqref}[1]{\mbox{Eq.~(\ref{#1})}}

\newcommand{\bra}[1]{\langle #1|}
\newcommand{\ket}[1]{|#1\rangle}

\newcommand{\abs}[1]{\left|#1\right|}

\newcommand{\be}{\begin{equation}}
\newcommand{\ee}{\end{equation}}
\newcommand{\bea}{\begin{eqnarray}}
\newcommand{\eea}{\end{eqnarray}}

\begin{document}

\title{Rate-equation approach for multi-level quantum systems
}

\author{M.~P.~Liul}
\email[e-mail:]{liul@ilt.kharkov.ua}
\affiliation{B.~Verkin Institute for Low Temperature Physics and Engineering, Kharkov 61103, Ukraine}

\author{S.~N.~Shevchenko}
\affiliation{B.~Verkin Institute for Low Temperature Physics and Engineering, Kharkov 61103, Ukraine}
\affiliation{V.~N.~Karazin Kharkiv National University, Kharkov 61022, Ukraine}

\date{\today}
\begin{abstract}
Strong driving of quantum systems opens opportunities for both controlling and characterizing their states. For theoretical studying of these systems properties we use the rate-equation formalism. The advantage of such approach is its relative simplicity. We used the formalism for description of a two-level system (TLS) with further expanding it on a case of a multi-level system. Obtained theoretical results have good agreement with experiments. The presented approach can also be considered as one more way to explore properties of quantum systems and underlying physical processes such as, for instance, Landau-Zener-St\"{u}ckelberg-Majorana transitions and interference. 

\end{abstract}

\maketitle

\section{Introduction}
Any problems related to quantum computers are very actual in modern physics [\onlinecite{Nielsen2010}, \onlinecite{Haffner2008}]. Superconducting qubits can be considered as very good candidates for being building blocks of these devices [\onlinecite{Devoret2013}, \onlinecite{Gambetta2017}, \onlinecite{Wendin2017}] since they have the following advantages~[\onlinecite{Oliver2013}]: it is possible to control superconducting qubits by microwaves; such systems show good performance during operations at nanosecond scales; superconducting qubits are scalable what opens opportunities to use them in lithography. 
\begin{figure*}
	\includegraphics[width=0.9 \linewidth]{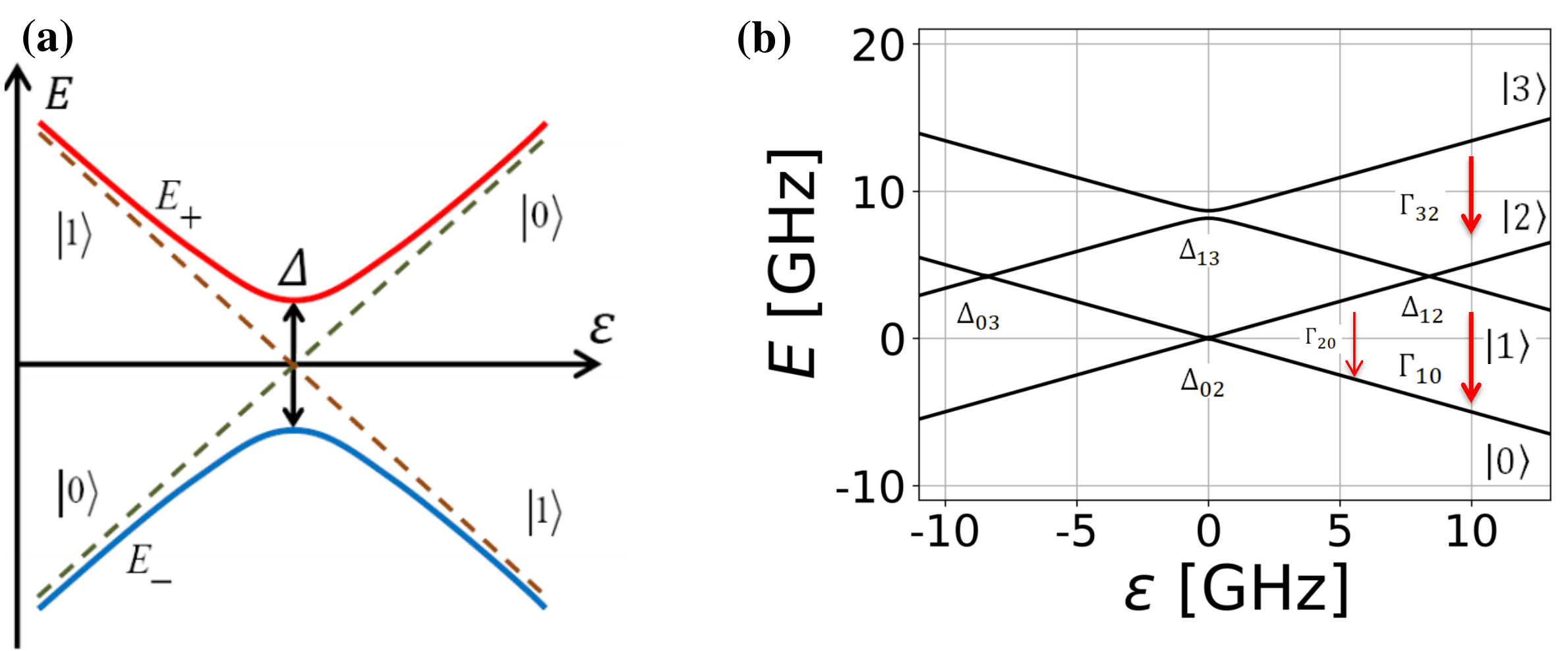}
	\caption{Energy levels as a function of the energy bias $\varepsilon$. Panel (a) shows graph for a TLS, (b) depicts energy levels of the four-level system studied in Ref.~[\onlinecite{Berns2008}].}
		\label{tls_energy_levels}
\end{figure*}

As a result we can conclude that any studying of superconducting qubits properties is very important for further growth and development of quantum computers. For example, such investigations could give useful insights for improvement of quantum logic gate operations [\onlinecite{Campbell2020}] and enhancement
of quantum algorithms performance in general [\onlinecite{Lucero2012}].

The presented research is also important because it gives one more approach for studying of the Landau-Zener-St\"{u}ckelberg-Majorana (LZSM) transitions and LZSM interferometry [\onlinecite{Oliver2005}, \onlinecite{Shevchenko2010}, \onlinecite{Sillanpaa2006}, \onlinecite{Kofman2022}]. LZSM transitions occur when a TLS is irradiated by a signal with the frequency which is much smaller than distance between energy levels [\onlinecite{Izmalkov2004}]. Such a phenomenon is reflected in various scientific fields such as nuclear physics [\onlinecite{Thiel1990}], quantum optics [\onlinecite{Bouwmeester1995}], chemical physics [\onlinecite{Zhu1997}], solid-state physics [\onlinecite{Wernsdorfer2000}], quantum information science [\onlinecite{Fuchs2011}]. Especially, it is possible to use such transitions for increasing tunneling rate [\onlinecite{Ankerhold2003}, \onlinecite{Ithier2005}], controlling qubit gate operations [\onlinecite{Saito2004}], preparing quantum states [\onlinecite{Saito2006}, \onlinecite{Ribeiro2009}], multi-signal spectroscopy [\onlinecite{Nakonechnyi2021}]. 

The repetition of LZSM transitions leads to LZSM interference [\onlinecite{Shevchenko}, \onlinecite{Ivakhnenko2022}]. The LZSM interferometry can be used for a system description and control, what was underlined in Refs.~[\onlinecite{Gorelik1998}, \onlinecite{Wu19}, \onlinecite{Ivakhnenko2022}]. LZSM interferometry allows to understand better the results of experiments which studied photon-assisted transport, conducted by periodic waves, in superconducting systems [\onlinecite{Tien1963}, \onlinecite{Nakamura1999}] and in quantum dots [\onlinecite{Kouwenhoven1994}, \onlinecite{Naber2006}]. The result of interaction of a quantum system with environment is decoherence. Such an effect is reflected in behavior of interference picture [\onlinecite{Berns2006}, \onlinecite{Rudner2008}, \onlinecite{Du2010}, \onlinecite{Malla2019}, \onlinecite{Malla2022}]. Thus, information about decoherence processes can be deduced from the LZSM interference picture. 

The rest of the paper is organized as follows. In Sec.~II the rate-equation formalism for TLS is introduced with its expansion on multi-level systems. Sec.~III is devoted to application of a considered approach to study stationary regime of a persistent current qubit, explored by authors of Ref.~[\onlinecite{Berns2006}]. The analysis of the persistent current qubit dynamics was implemented in Sec.~IV. In Sec.~V we adopt the rate-equation formalism for describing a multi-level system, proposed in Ref.~[\onlinecite{Berns2008}]. It is noticeable to mention that theoretical and experimental results are in very good agreement. In Sec.~VI we make conclusions. 
\section{Rate-equation approach: from two-level systems to multi-level systems}
The authors of Refs.~[\onlinecite{Berns2006}, \onlinecite{Oliver2009}] successfully described their experiment within the rate-equation formalism (see also Refs.~[\onlinecite{Ferron2010}, \onlinecite{Ferron2012}, \onlinecite{Ferron2016}]). In this section we give a short description of theoretical aspects of this method.

Let us firstly employ this method for a TLS with following extension of obtained results on multi-level systems. The Hamiltonian of a TLS, driven by external field can be written in the form:
\begin{eqnarray}
 \widehat{H}(t) = -\frac{\mathrm{\Delta}}{2}\widehat{\sigma}_{x} - \frac{h(t)}{2}\widehat{\sigma}_{z}, 
\label{eq:Hamiltonian_TLS}
\end{eqnarray}
where $\widehat{\sigma}_{z} = \begin{pmatrix} 1 & 0 \\ 0 & -1 \end{pmatrix}$ and $\widehat{\sigma}_{x} = \begin{pmatrix} 0 & 1 \\ 1 & 0 \end{pmatrix}$ are Pauli matrices, $\mathrm{\Delta}$ is the level splitting, $h(t)$ is the external excitation which can be presented as follows:
\begin{eqnarray}
 h(t) = \varepsilon + A\sin2\pi\nu t + \delta \varepsilon_{\mathrm{noise}}(t).
\label{eq:excitation}
\end{eqnarray}
Here $\varepsilon$ is an energy detuning, $\nu$ and $A$ are the frequency of the excitation field and its amplitude respectively, $\delta\varepsilon_{\mathrm{noise}}(t)$ can be treated as the classical noise. In paper [\onlinecite{Berns2006}] the authors used white-noise model and for the LZSM transition rate they obtained (see also Refs.~[\onlinecite{Chen2011}, \onlinecite{Wang2010}, \onlinecite{Wen2010}, \onlinecite{Otxoa2019}])
\begin{eqnarray}
W(\varepsilon, A) = \frac{\mathrm{\Delta}^{2}}{2}\sum_{n}\frac{\mathrm{\Gamma}_{2}J_{n}^{2}(A/\nu)}{\left ( \varepsilon - n \nu \right )^{2} + \mathrm{\Gamma}_{2}^{2}}.
\label{eq:transition_rate}
\end{eqnarray}
Here $\mathrm{\Gamma}_{2}$ is the decoherence rate, $J_{n}$ is the Bessel function, and the reduced Planck constant is equal to unity ($\hbar~=~1$). The diagram of TLS energy levels is depicted in Fig.~\ref{tls_energy_levels}(a). Eq.~(\ref{eq:transition_rate}) characterizes the transitions which happen when a system passes through a point of maximum levels convergence.
\begin{figure*}
	\includegraphics[width=1 \linewidth]{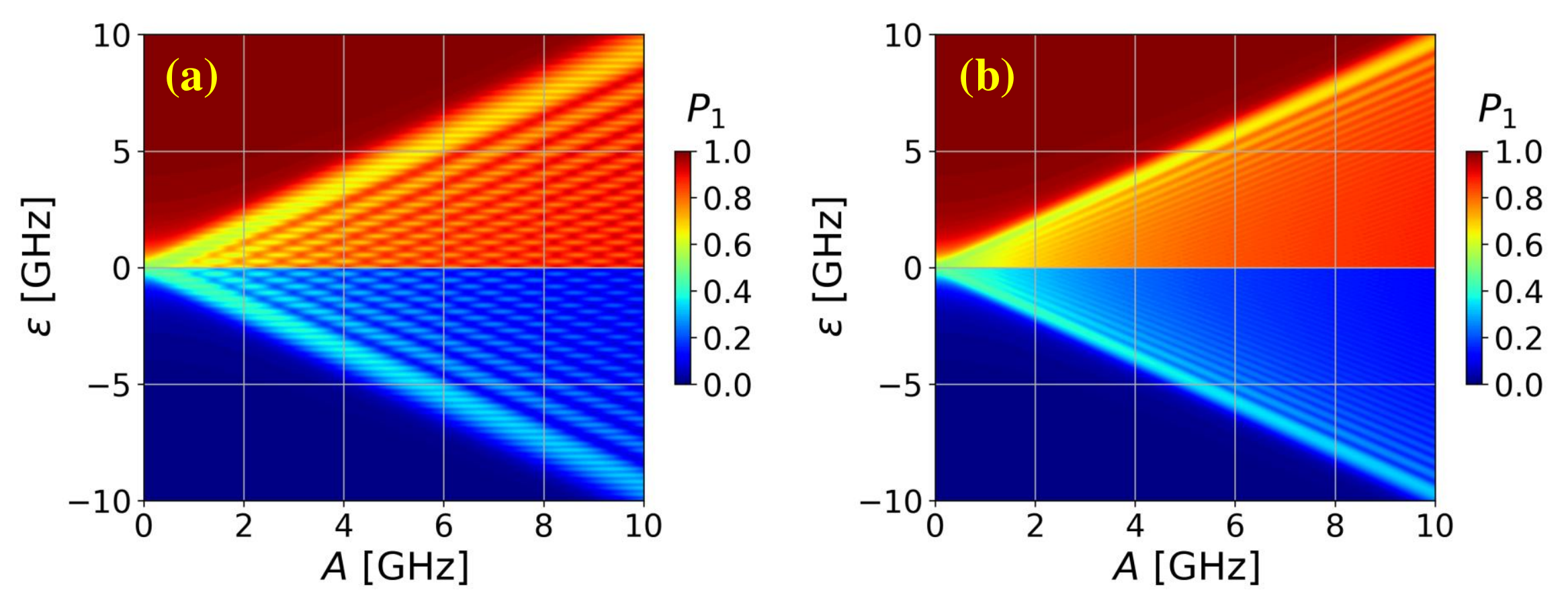}
	\caption{Population $P_{1}$ as a function of the excitation field amplitude $A$ and the energy detuning $\varepsilon$. The computations were done for two different values of the excitation field frequency: (a)~$\nu = \unit[270]{MHz}$ and (b)~$\nu = \unit[90]{MHz}$, in agreement with Ref.~[\onlinecite{Berns2006}]. The used parameters are $\mathrm{\Delta}~=~\unit[13]{MHz}$, $\mathrm{\Gamma}_1~=~\unit[50]{kHz}$, $\mathrm{\Gamma}_2~=~\unit[95]{MHz}$, $\mathrm{\Gamma}'_1~=~\mathrm{\Gamma}_1\times 10^{-3}$.}
		\label{tls_theory}
\end{figure*}
In the case of a multi-level system we should assign a corresponding transition rate to each level quasicrossing point (point of maximum levels convergence). 
The authors of Ref.~[\onlinecite{Wen2009}] proposed to extend the Eq.~(\ref{eq:transition_rate}) on the transition between arbitrary states $\ket{i}$ and $\ket{j}$ of a multi-level system by the formula: 
\begin{eqnarray}
W_{ij}(\varepsilon_{ij}, A) = \frac{\mathrm{\Delta}_{ij}^{2}}{2}\sum_{n}\frac{\mathrm{\Gamma}_{2}J_{n}^{2}(A/\nu)}{\left ( \varepsilon_{ij} - n \nu \right )^{2} + \mathrm{\Gamma}_{2}^{2}},
\label{eq:transition_rate_general}
\end{eqnarray}
 where $\mathrm{\Delta}_{ij}$ is the energy splitting between states $\ket{i}$ and $\ket{j}$, $\varepsilon_{ij}$ is the corresponding energy detuning. Then the rate equation for the $\ket{i}$ state can be expressed
\begin{eqnarray}
\frac{d P_{i}}{dt} = \sum_{j}W_{ij}(P_{j} - P_{i}) + \sum_{i'}\mathrm{\Gamma}_{i' i}P_{i'} - \sum_{i'}\mathrm{\Gamma}_{i i'}P_{i}. 
\label{eq:rate_equation_general}
\end{eqnarray}
Here $P_{i}$ is the probability that a system occupies $\ket{i}$ state, $\mathrm{\Gamma}_{ii'}$ characterize the relaxation from the state $\ket{i}$ to the state $\ket{i'}$. \\

Thus, writing equations (\ref{eq:rate_equation_general}) for each level we can find occupation probabilities of the levels and then build corresponding interferograms. Usually for simplicity one considers only a stationary case, $d P_{i}/dt = 0$. The solution of such a system will not describe a quantum object dynamics, but it is suitable for obtaining its main properties. Also we can use the fact that the sum of all probabilities is equal to unity $\sum_{i}P_{i} = 1$. 

\section{Qubit: interferogram}

We start the studying of the rate-equation formalism from applying it to a two-level system, proposed in Ref.~[\onlinecite{Berns2006}]. The considered system is a persistent-current qubit [\onlinecite{Orlando1999}] described by the Hamiltonian from Eq.~(\ref{eq:Hamiltonian_TLS}). The rate equation (\ref{eq:rate_equation_general}) for the system can be rewritten in the form:
\begin{eqnarray}
\frac{d P_{1}}{dt} = W_{10}(P_{0} - P_{1}) + \mathrm{\Gamma}'_{1}P_{0} - \mathrm{\Gamma}_{1}P_{1}, 
\label{eq:rate_equation_two_level}
\end{eqnarray}
where $\mathrm{\Gamma}_{1}$ is the relaxation rate from the state $\ket{1}$ to the state $\ket{0}$, $\mathrm{\Gamma}'_{1}$ characterizes the relaxation from the state $\ket{0}$ to the state $\ket{1}$. Since we are interested in the stationary regime we can put $d P_{i}/dt = 0$. Supplementing Eq.~(\ref{eq:rate_equation_two_level}) by the relation $P_{0} + P_{1} = 1$ we find:
\begin{eqnarray}
P_{0} = \frac{W_{10} + \mathrm{\Gamma}_{1}}{2W_{10} + \mathrm{\Gamma}_{1} + \mathrm{\Gamma}'_{1}}, ~~~ 
P_{1} = \frac{W_{10} + \mathrm{\Gamma}'_{1}}{2W_{10} + \mathrm{\Gamma}_{1} + \mathrm{\Gamma}'_{1}}.
\label{eq:system_six_5}
\end{eqnarray}

In Ref.~[\onlinecite{Berns2006}] the occupation probability of an upper charge state $\ket{1}$ $P_{1}$ as a function of the flux detuning $\mathrm{\Delta} f$ (the energy detuning $\varepsilon$) and the source voltage $V_{\mathrm{rms}}$ (amplitude of the excitation field $A$ in theory) were experimentally studied for two values of the excitation field frequency: (a)~$\nu~=~\unit[270]{MHz}$ and (b)~$\nu~=~\unit[90]{MHz}$. The corresponding plot is shown in Fig.~2 of Ref.~[\onlinecite{Berns2006}]. The parameters of the experiment are $\mathrm{\Delta}~=~\unit[13]{MHz}$, $\mathrm{\Gamma}_1~=~\unit[50]{kHz}$, $\mathrm{\Gamma}_2~=~\unit[95]{MHz}$, $\mathrm{\Gamma}'_1~= ~\mathrm{\Gamma}_1\exp{(-\beta \varepsilon)}$, where $\beta$ is a parameter which describes the relaxation from the lower level to the upper one. For our theoretical calculations we assumed $\mathrm{\Gamma}'_1~=~\mathrm{\Gamma}_1\times 10^{-3}$. The results of theoretical computations are presented in Fig.~\ref{tls_theory}. We can conclude that theoretical and experimental plots are in a good agreement.

\section{Qubit: dynamics}
\begin{figure*}
	\includegraphics[width=1 \linewidth]{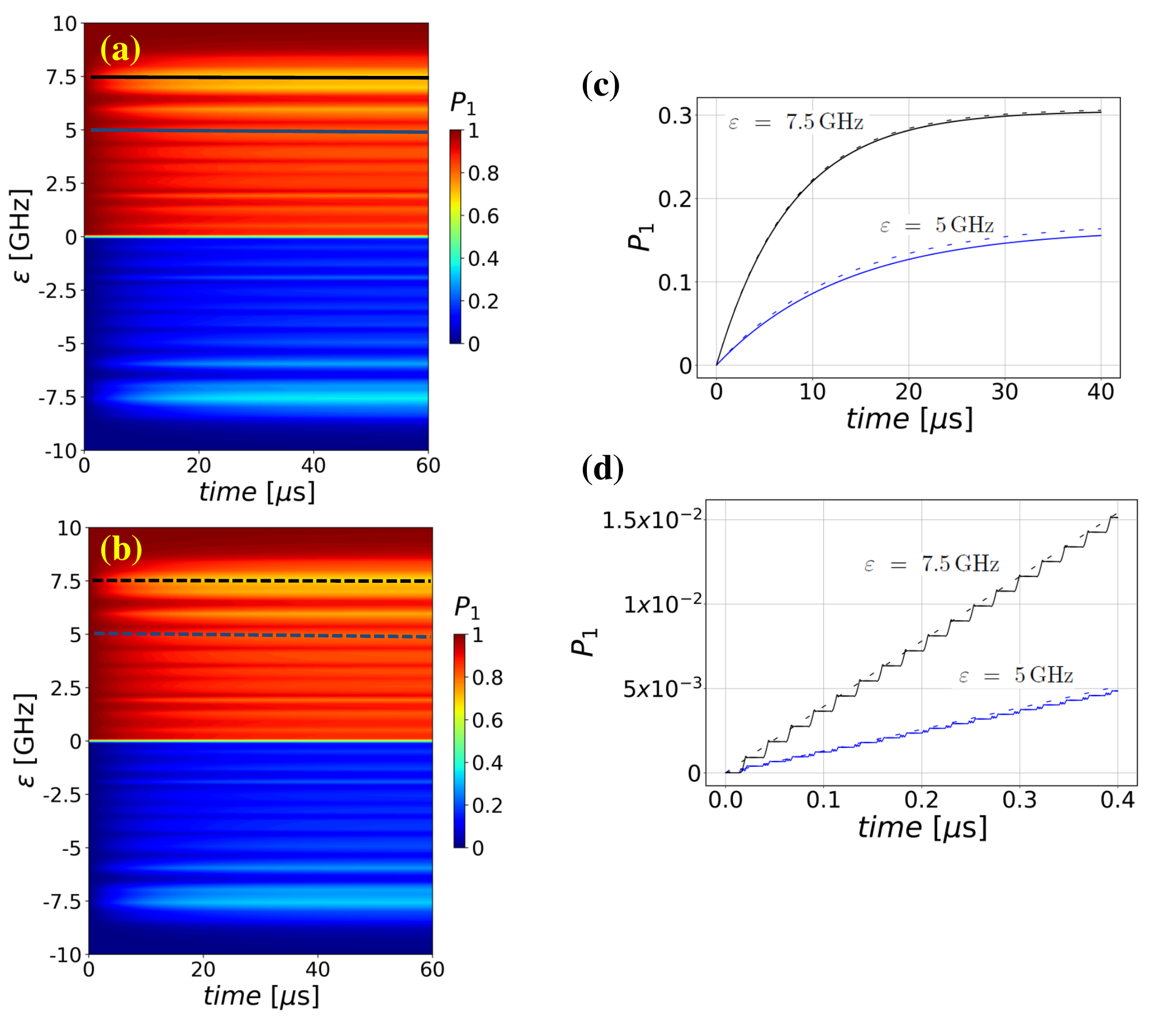}
	\caption{Population $P_{1}$ as a function of time $t$. Panels (a, b) show $P_{1}$ as a function of time $t$ and energy detuning $\varepsilon$ for $A~=~\unit[8]{GHz}$ and $\nu~=~\unit[270]{MHz}$. Panel~(a)~was calculated by the Lindblad equation approach, (b)~is the result of solving the rate equations. Panel~(c) is the line cut along (a, b) at  $\varepsilon~=~\unit[5]{GHz}$ (blue line, lower set of curves) and $\varepsilon~=~\unit[7.5]{GHz}$ (black line, upper set of curves). Solid lines correspond to the exact solution, dashed are solutions of the rate equations. Panel (d) shows the dynamics of the considered process during the first 400 nanoseconds, the lines are marked in the same way as in (c). Other parameters are the same with Fig.~\ref{tls_theory}.} 
		\label{tls_time}
\end{figure*}

In this section a qubit dynamics is considered. For the analysis we compare two approaches: solving of the Lindblad equation (the exact solution) and the system of rate equations (the approximate one). Let us firstly describe the exact approach (see, for example, Refs.~[\onlinecite{Lindblad1976}, \onlinecite{Manzano2020}]). The Lindblad equation with the Hamiltonian (\ref{eq:Hamiltonian_TLS}) can be written in the form:
\begin{eqnarray}
\frac{d\rho}{d t} = -i\left [ \widehat{H}, \rho \right ] + \sum_{\alpha}\breve{L}_{\alpha}\left [ \rho \right ],
\label{Bloch_eq}
\end{eqnarray}
where $\rho = \begin{pmatrix} \rho_{00} & \rho_{01} \\ \rho^{*}_{01} & 1 - \rho_{00} \end{pmatrix}$ is the density matrix, such that $P_{1} = 1 - \rho_{00}$, $\breve{L}_{\alpha}$ is the Lindblad superoperator, which describes the relaxation of the system caused by interaction with the environment,
\begin{eqnarray}
\breve{L}_{\alpha}\left [ \rho \right ] = \widehat{L}_{\alpha}\rho \widehat{L}_{\alpha}^{+} - \frac{1}{2}\left\{\widehat{L}_{\alpha}^{+}\widehat{L}_{\alpha}, \rho \right\},
\label{Lindblad_superoperator}
\end{eqnarray}
where $\left\{a,b \right\} = ab + ba$ is the anticommutator. For a qubit there are two possible channels of relaxation: dephasing (described by $\widehat{L}_{\phi}$) and energy relaxation (described by $L_{\mathrm{relax}}$). The corresponding operators have the following form:%
\begin{eqnarray}
\widehat{L}_{\mathrm{relax}} = \sqrt{\mathrm{\Gamma}_1}\widehat{\sigma}^{+}, ~ ~
\widehat{L}_{\phi} = \sqrt{\frac{\mathrm{\Gamma}_{\phi}}{2}}\widehat{\sigma}_{z},
\end{eqnarray}
where $\widehat{\sigma}^{+} = \begin{pmatrix} 0 & 1 \\ 0 & 0 \end{pmatrix}$, $\mathrm{\Gamma}_1$ is the qubit relaxation, $\mathrm{\Gamma}_{\phi}$ is the pure dephasing rate, $\mathrm{\Gamma}_2 = \mathrm{\Gamma}_1/2 + \mathrm{\Gamma}_{\phi}$ is the decoherence rate.

On the one hand, by solving Eq.~(\ref{Bloch_eq}) one obtains $P_{1}$ as a function of time $t$, driving frequency $\nu$ and amplitude $A$, energy detuning $\varepsilon$, the level splitting $\mathrm{\Delta}$. The occupation probability is the function of all these parameters, $P_{1} = P_{1}(t,~ \nu,~ A,~ \varepsilon,~\mathrm{\Delta})$. Obtained dependence allows us to build, for instance, $P_{1} = P_{1}(\varepsilon, t)$. On the other hand, we can get the same relation by solving Eq.~(\ref{eq:rate_equation_general}). Figs.~\ref{tls_time}~(a, b) show the results of the theoretical calculations of $P_{1}$ as a function of time $t$ and energy detuning $\varepsilon$ for $A~=~\unit[8]{GHz}$ and $\nu~=~\unit[270]{MHz}$, other parameters are the same with Fig.~\ref{tls_theory}. Panel (a) was calculated by the Lindblad equation approach, while (b) is the result of solving the rate equations. One can conclude that considered approaches are in a good qualitative correspondence. We also built the pictures for the $\nu~=~\unit[90]{MHz}$ but since it did not give any additional insights, it was decided not to include this case to the article. 
In Fig.~\ref{tls_time}~(c), we can see the line cut along Figs.~\ref{tls_time}~(a, b) at  $\varepsilon~=~\unit[5]{GHz}$ (blue line) and $\varepsilon~=~\unit[7.5]{GHz}$ (black line). Solid lines correspond to the exact solution, dashed lines are solutions of the rate equations. We can see that both approaches are in a good agreement. The difference between them can be seen if to zoom pictures (for example, consider the first microseconds of the process). Fig.~\ref{tls_time}~(d) shows the dynamics of the considered process during the first 400 nanoseconds. The lines are marked in the same way as in Fig.~\ref{tls_time}~(c). From the comparison we can deduce that the rate-equation formalism averages the oscillations, so the corresponding curve is a monotonous curve, while the Lindblad equation approach reflects more sophisticated system behavior. 
\begin{figure*}
	\includegraphics[width=1 \linewidth]{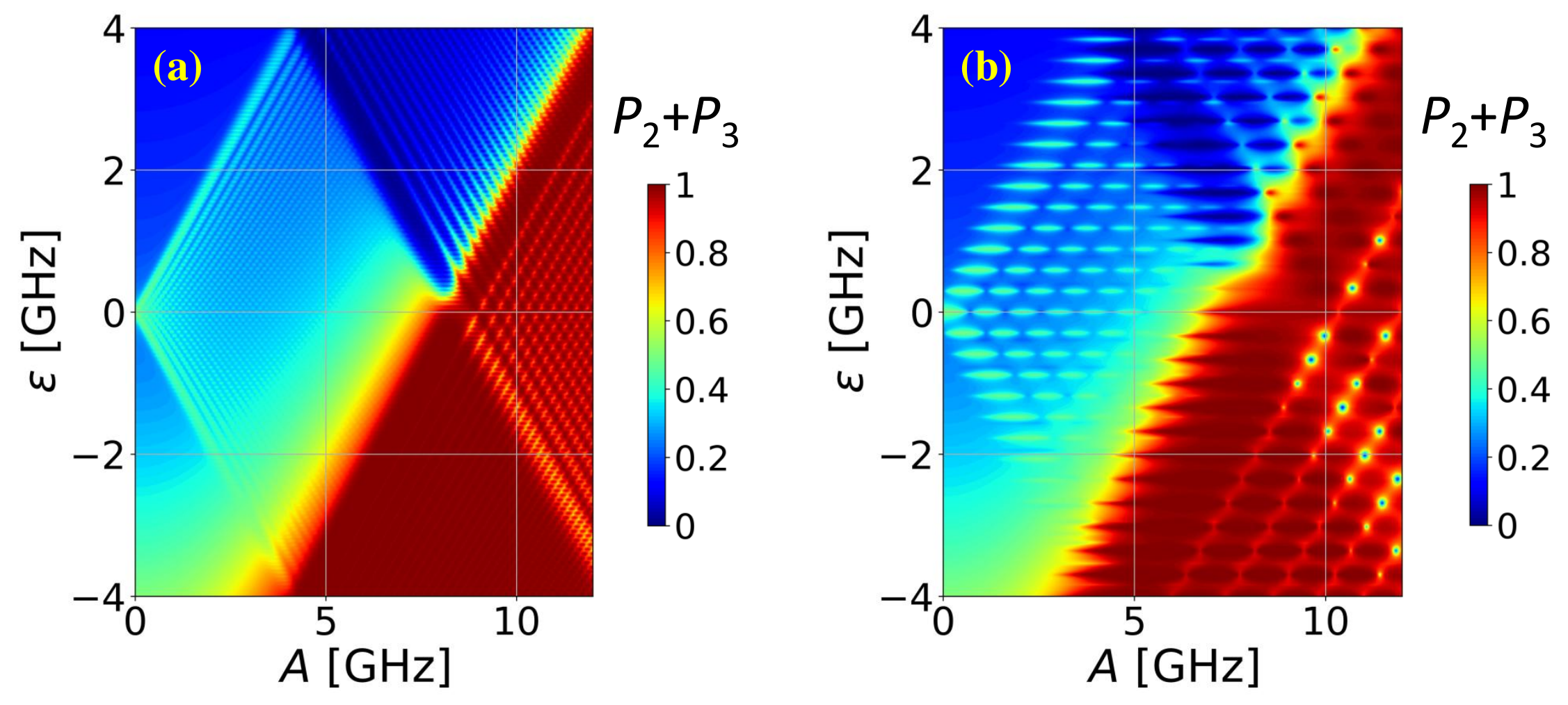}
	\caption{Artificial atom population $P_{\mathrm{L}} = P_{2} + P_{3}$ as a function of the excitation field amplitude $A$ and the energy detuning $\varepsilon$. For panel (a) the driving frequency $\nu~=~\unit[0.16]{GHz}$, for (b) $\nu~=~\unit[0.85]{GHz}$. The corresponding relaxation rates of the system are $\mathrm{\Gamma}_{10}~=~\unit[0.6]{GHz}$, $\mathrm{\Gamma}_{32}~=~\unit[0.6]{GHz}$, $\mathrm{\Gamma}_{20}~=~\unit[0.05]{MHz}$. The inverse relaxation rates (from a lower state $\ket{m}$ to an upper one $\ket{n}$) are Boltzmann suppressed and for simplicity we took $\mathrm{\Gamma}_{mn}~=~\mathrm{\Gamma}_{nm}/100$. The energy splittings are equal $\mathrm{\Delta}_{02}~=~\unit[0.09]{GHz}$, $\mathrm{\Delta}_{12}~=~\unit[0.013]{GHz}$, $\mathrm{\Delta}_{13}~=~\unit[0.5]{GHz}$, $\mathrm{\Delta}_{03}~=~\unit[0.5]{GHz}$ and their positions are at $\varepsilon~=~\unit[0]{},~\unit[8.4],~\unit[0]{}$ and $~\unit[-8.4]{GHz}$ respectively. The decoherence rate $\mathrm{\Gamma}_{2}~=~\unit[0.05]{GHz}$. The system energy slopes [\onlinecite{Berns2008}] equal $\abs{m_{0}}$, $\abs{m_{2}}~=~\unit[1.44]{}$, $\abs{m_{1}}$, $\abs{m_{3}}~=~\unit[1.09]{}$.} 
		\label{four_level_theory}
\end{figure*}
\begin{figure*}
	\includegraphics[width=0.4 \linewidth]{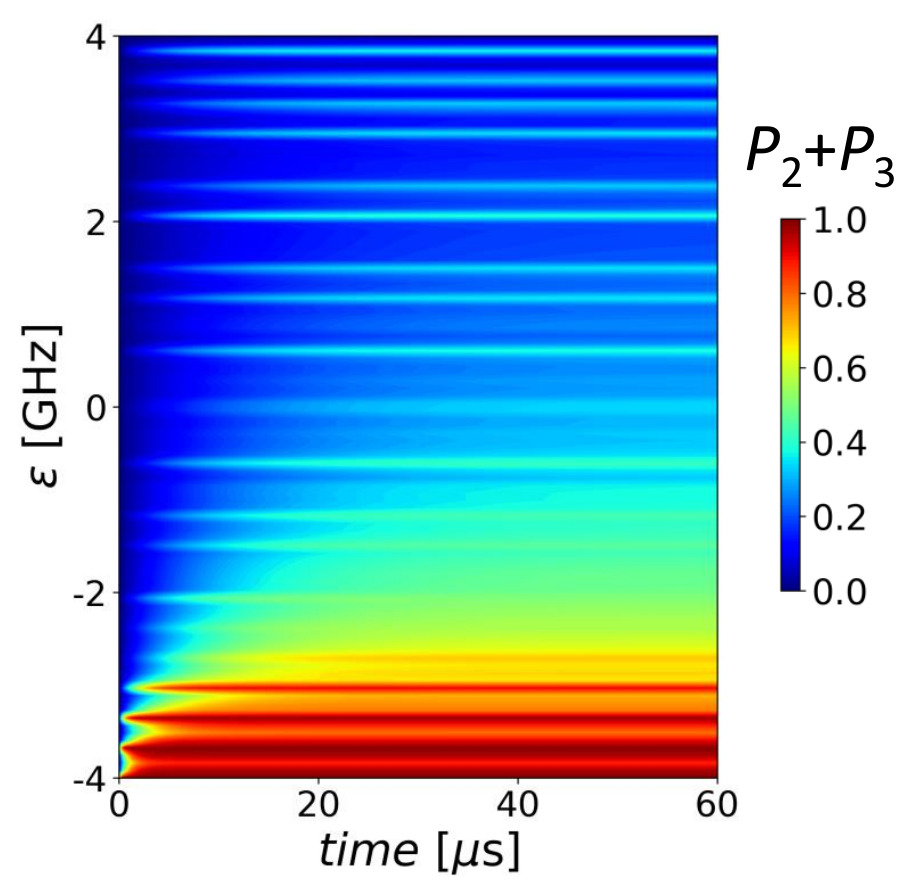}
	\caption{Artificial atom population $P_{\mathrm{L}} = P_{2} + P_{3}$ as a function of the energy detuning $\varepsilon$ and time $t$. The calculations were done for $A~=~\unit[4]{GHz}$ and $\nu~=~\unit[0.85]{GHz}$. Other parameters are the same with Fig.~\ref{four_level_theory}.} 
		\label{four_level_theory_time}
\end{figure*}
\section{Multi-level systems: dynamics and interferograms; the case of a solid-state artificial atom}

In this section we theoretically study the solid-state artificial atom in the layout of Ref Ref.~[\onlinecite{Berns2008}]. An artificial atom is a structure where electrons are trapped and can only have discrete energy states, like in real atoms. The unperturbed part of the considered system Hamiltonian has a form [\onlinecite{Whiticar2022}]: 
\begin{eqnarray}
H = \sum_{n}E_{n}\ket{n}\bra{n} - \frac{1}{2}\sum_{m\neq n}\mathrm{\Delta}_{mn}\ket{m}\bra{n},
\label{Hamiltonian_4_level_general}
\end{eqnarray}
for case of our system unperturbed part of the Hamiltonian can be written as
\begin{eqnarray}
H = -\frac{1}{2}\begin{pmatrix}
-\varepsilon - B &  0&\mathrm{\Delta}_{02}  & \mathrm{\Delta}_{03}\\ 
 0& \varepsilon & \mathrm{\Delta}_{12} &\mathrm{\Delta}_{13} \\ 
 \mathrm{\Delta}_{02}& \mathrm{\Delta}_{12} & -\varepsilon & 0\\ 
 \mathrm{\Delta}_{03}& \mathrm{\Delta}_{13} & 0 & \varepsilon - B
\end{pmatrix}.
\label{Hamiltonian_4_level_matrix}
\end{eqnarray}
The value $B~=~~\unit[2 \times 8.4]{GHz}$ describes the position of quasicrossings $\mathrm{\Delta}_{12}$ and $\mathrm{\Delta}_{03}$ (see also further in the text). The corresponding energy diagram can be seen in  Fig.~\ref{tls_energy_levels}(b). The obtained energy diagram is in a good agreement with ones in Refs.~[\onlinecite{Berns2008}, \onlinecite{Wen2009}]. In the region of our interest the system contains 4 energy levels, placed in double-well potential, detailed energy configuration can be found in Ref.~[\onlinecite{Wen2010}]. In the considered case states $\ket{0}$ and $\ket{1}$ are in the right well, states $\ket{2}$ and $\ket{3}$ are in the left one. Moreover, accordingly to Ref.~[\onlinecite{Berns2008}], the relaxation inside a well is faster in this solid-state artificial atom than the relaxation between wells, so one can neglect relaxations from state $\ket{1}$ to state $\ket{2}$ and vise-versa. In the experiment the population in the left well $P_{\mathrm{L}} = P_{2} + P_{3}$ was measured. Applying Eq.~(\ref{eq:rate_equation_general}) to the analyzed system one obtains the system of rate equations:

\begin{align}
\resizebox{0.5\textwidth}{!}{$
\begin{cases}
\dot{P_{0}} = -P_{0}(W_{02} + W_{03} + \mathrm{\Gamma}_{20}) + P_{2}(W_{20} + \mathrm{\Gamma}_{20})+ P_1\mathrm{\Gamma}_{10} + P_{3}W_{03} \\ 
\dot{P_{1}} = -P_{1}(W_{12} + W_{13} + \mathrm{\Gamma}_{10}) + P_{2}W_{12} + P_{3}W_{13} \\ 
\dot{P_{2}} = P_{0}(W_{02} + \mathrm{\Gamma}_{02})  - P_{2}(W_{02} + W_{12} + \mathrm{\Gamma}_{20})+  P_{3}\mathrm{\Gamma}_{23} + P_{1}W_{12} \\ 
P_{0} + P_{1} + P_{2} + P_{3} = 1.\\
\end{cases}$}
\label{eq:four_level_system}
\end{align}
The corresponding relaxation rates of the system are $\mathrm{\Gamma}_{10}~=~\unit[0.6]{GHz}$, $\mathrm{\Gamma}_{32}~=~\unit[0.6]{GHz}$, $\mathrm{\Gamma}_{20}~=~\unit[0.05]{MHz}$. The inverse relaxation rates (from a lower state $\ket{m}$ to an upper one $\ket{n}$) are Boltzmann suppressed and for simplicity we took $\mathrm{\Gamma}_{mn}~=~\mathrm{\Gamma}_{nm}/100$. The energy splittings are equal to $\mathrm{\Delta}_{02}~=~\unit[0.09]{GHz}$, $\mathrm{\Delta}_{12}~=~\unit[0.013]{GHz}$, $\mathrm{\Delta}_{13}~=~\unit[0.5]{GHz}$, $\mathrm{\Delta}_{03}~=~\unit[0.5]{GHz}$ and their positions are at $\varepsilon~=~\unit[0]{},~\unit[8.4],~\unit[0]{}$ and $~\unit[-8.4]{GHz}$ respectively. The decoherence rate $\mathrm{\Gamma}_{2}~=~\unit[0.05]{GHz}$.

To make the correspondence between the theory and the experiment better, the authors of Ref.~[\onlinecite{Wen2009}] proposed to take into account the diabatic energy-level slope $m_{i}=dE_{i}(\varepsilon)/d\varepsilon$ of a level $i$ with energy $E_{i}$. The Eq.~(\ref{eq:excitation}) can be rewritten in the form:
\begin{eqnarray}
 h_{ij}(t) = (\abs{m_{i}}+\abs{m_{j}})(\varepsilon + A\sin2\pi\nu t) + \delta \varepsilon_{\mathrm{noise}}(t).
\label{eq:excitation_general}
\end{eqnarray}
The system energy slopes [\onlinecite{Berns2008}] equal $\abs{m_{0}}$, $\abs{m_{2}}~=~\unit[1.44]{}$, $\abs{m_{1}}$, $\abs{m_{3}}~=~\unit[1.09]{}$. 

The results of the theoretical calculations are presented in Fig.~\ref{four_level_theory}. Panel (a) corresponds to the case of $\nu~=~\unit[0.16]{GHz}$, (b) was built for driving frequency $\nu~=~\unit[0.85]{GHz}$. The full picture consists of triangles which can be very roughly interpreted as interactions within TLS. For example, the system behaves like a qubit on the interval $A~<~\unit[4]{GHz}$. For the case (a) the picture on the interval $A~>~\unit[8.4]{GHz}$ is also TLS-like, while for the case (b) the system behavior is more sophisticated. We can also conclude that for the higher frequency resonances become more distinguishable as it was observed for a qubit. 

To complete the research let us study the system dynamics. Fig.~\ref{four_level_theory_time} shows a dependence of population $P_{2} + P_{3}$ on time and energy detuning $\varepsilon$ for $A~=~\unit[4]{GHz}$ and $\nu~=~\unit[0.85]{GHz}$. All parameters are the same with Fig.~\ref{four_level_theory}. 

\section{Conclusions}
Description of an $N$-level quantum system, if solving a Master equation, requires solving $N^{2}-1$ equations for the density-matrix components. We consider an alternative approach consisting in solving the rate equations, the number of which is $N-1$. We started from a TLS for which we have only one equation instead of three Bloch equations. Then we considered generalization for a multi-level system and described a multi-level flux-qubit-based device. The rate-equation approach involves relaxation and decoherence and is demonstrated to be convenient for obtaining the stationary states. Particularly, we have applied this method for the
LZSM interferometry, which is an important tool for quantum characterization and control.

\begin{acknowledgments}

The authors acknowledge fruitful discussions with A.~Ryzhov. This work was supported by Army Research Office (ARO) (Grant No.~W911NF2010261). 

\end{acknowledgments}

\nocite{apsrev41Control} 
\bibliographystyle{apsrev4-1}
\bibliography{references}

\end{document}